\documentclass[conference]{IEEEtran}
\IEEEoverridecommandlockouts

\usepackage{balance}
\usepackage[cmex10]{amsmath}
\usepackage{amssymb}
\usepackage{ifpdf}
\usepackage{cite}
\usepackage{hyperref}

\ifCLASSINFOpdf
   \usepackage[pdftex]{graphicx}
   \graphicspath{{./pdf/}}
   \DeclareGraphicsExtensions{{.pdf}}
\else
   \usepackage[dvips]{graphicx}
   \graphicspath{{./eps/}}
   \DeclareGraphicsExtensions{{.eps}}
\fi

\newcommand{\comment}[2]{#2}

\usepackage{multirow}

\begin{document}
\hyphenation{multi-symbol}
\title{ Closing the Gap to the Capacity of APSK:
Constellation Shaping and Degree Distributions
}
\author{ Xingyu Xiang and Matthew C. Valenti\\
Lane Department of Computer Science and Electrical Engineering\\
West Virginia University, Morgantown, WV, U.S.A.
\thanks{The authors were sponsored by the National Science Foundation under Award No. CNS-0750821 and by the United States Army Research Laboratory under Contract W911NF-10-0109.}}
\date{}

\maketitle

\thispagestyle{empty}

\begin{abstract}
Constellation shaping is an energy-efficient strategy involving the transmission of lower-energy signals more frequently than higher-energy signals.   Previous work has shown that shaping is particularly effective when used with coded amplitude phase-shift keying (APSK), a modulation that has been popularized recently due to its inclusion in the DVB-S2 standard.   While shaped APSK can provide significant gains when used with standard off-the-shelf LDPC codes, such as the codes in the DVB-S2 standard, additional non-negligible gains can be achieved by optimizing the LDPC code with respect to the shaped APSK modulation.
In this paper, we optimize the degree distributions of the LDPC code used in conjunction with shaped APSK.  The optimization process is an extension of the EXIT-chart technique of ten Brink, et al., which has been adapted to account for the shaped APSK modulation.  We begin by constraining the code to have the same number of distinct variable-node degrees as the codes in the DVB-S2 standard, and show that the optimization provides 32-APSK systems with an additional coding gain of 0.34 dB at a system rate of $R=3$ bits per symbol, compared to shaped systems that use the long LDPC code from the DVB-S2 standard.  We then increase the number of allowed variable node degrees by one, and find that an additional 0.1 dB gain is achievable.
\end{abstract}

\section{Introduction}
The combination of low-density parity-check (LDPC) codes and amplitude phase-shift keying (APSK) has become popularized as a result of the widespread adoption of the Digital Video Broadcasting Satellite - Second Generation (DVB-S2) standard, which uses LDPC-coded APSK for the satellite delivery of high-definition television, interactive services, and data-content distribution \cite{dvb:2009}.  
An APSK signal constellation consists of several concentric rings of symbols, with each ring having a different amplitude and containing symbols that are separated by a constant phase offset \cite{gaudenzi:2006}.  When combined with LDPC codes, APSK offers an attractive combination of spectral and energy efficiency, and is particularly well suited for the nonlinearities in satellite channels that arise due to the use of high-power amplifiers.

\begin{figure*}[th]
\centering
\includegraphics[width=6in]{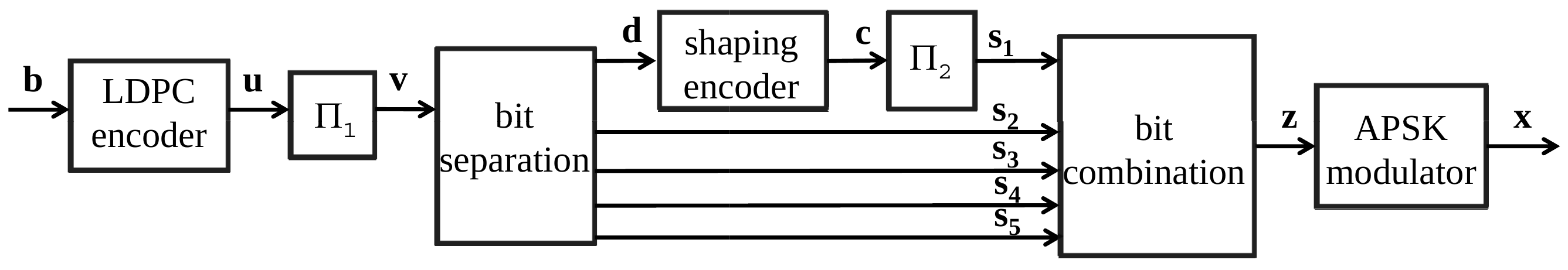}
\caption{Transmitter structure.} \label{Fig_transmitter}
\vspace{-0.5cm}
\end{figure*}

In typical implementations, such as DVB-S2, the APSK symbols are selected with uniform probability. However, as we have shown in \cite{valenti:icc2011,valenti:2012itc}, the performance of APSK may be significantly improved by selecting lower-energy signals more frequently than higher-energy signals.  The strategy in \cite{valenti:icc2011,valenti:2012itc} \comment{\cite{valenti:icc2011,xiang:milcom2011,valenti:2012itc}} is an instance of the general concept of {\em constellation shaping} \cite{calderbank:1990,legoff:2007}. Our specific strategy for shaping APSK is adapted from a strategy originally proposed by Le Goff, et al., in \cite{legoff:2007} for bit-interleaved turbo-coded pulse-amplitude modulation (PAM).  The strategy partitions the base signal constellation into two or more sub-constellations of increasing average energy and uses a {\em shaping code} \cite{calderbank:1990} to select signals from the lower energy sub-constellations more often than the signals from the higher energy sub-constellations.  While shaping increases the per-iteration complexity of the system,  the error-rate performance can often be improved enough that fewer iterations are required compared to a uniform system operating at the same signal-to-noise ratio (SNR) \cite{valenti:2012itc}.

Shaping an APSK constellation can achieve significant gains, even if an ``off-the-shelf'' standard channel code is used.  For instance, in \cite{valenti:icc2011} we show a gain in excess of 0.6 dB when shaped APSK is combined with UMTS standardized turbo code at a rate of 3.85 bits/symbol, while in \cite{valenti:2012itc} \comment{\cite{xiang:milcom2011}} shaping gains in excess of 0.7 dB was observed for LDPC-coded 32-APSK operating at a rate of 3 bits/symbol.
While using standardized codes is convenient, a significant additional gain can be achieved by tailoring the code to match the characteristics of the shaped modulation.  This fact can be visualized through the use of extrinsic information transfer (EXIT) charts \cite{brink:2004}.  The EXIT chart of an LDPC-coded system comprises a pair of curves, which need be properly matched for optimal performance. Since shaping changes the corresponding EXIT curve, a code whose EXIT curves are well matched in a uniform modulation scheme may not yield the optimal performance for a shaped modulation system. By changing the variable-node degree distribution, it is possible to identify a code for the shaped system with better matched EXIT curves. As we show, the improvement in performance can be significant.

The remainder of this paper is organized as follows.  Section \ref{Sec_SystemModel} provides a model for bit-interleaved LDPC-coded APSK with constellation shaping.  Section \ref{dvbs2} compares the performance of a shaped system against that of a comparable uniform system operating at the same rate, with both systems using codes from the DVB-S2 standard.  Section \ref{ldpcopt} uses EXIT charts \cite{brink:2004} to identify better degree distributions for the uniform APSK system, while section \ref{ldpcoptshape} \comment{further adapts} extends the technique to find optimal degree distributions for the shaped system.  In these sections, the optimal degree distributions are given along with simulation results for LDPC-coded 32-APSK operating over an additive white Gaussian noise (AWGN) channel at a rate of 3 bits/symbol with LDPC codes of two lengths.  Finally, the paper concludes in Section \ref{Sec_Conclusion}.

\section{System Model} \label{Sec_SystemModel}

The system's transmitter is shown in Fig. \ref{Fig_transmitter}.  A length-$k_c$ vector $\mathbf b$ of information bits is fed into a rate $R_c = k_c/n_c$ LDPC encoder, which produces a length-$n_c$ codeword $\mathbf u$. The first interleaver $\Pi_1$ produces a vector $\mathbf v$ after random permutation.  A {\em bit separator} separates $\mathbf v$ into $m$ streams, where $m = \log_2 M$ and $M$ is the number of symbols in the APSK constellation.  For the remainder of this paper, we assume that $M=32$, and hence, $m=5$.   The first stream, denoted $\mathbf d$, is passed through a rate $R_s = k_s/n_s$ shaping encoder to produce the shaped codeword $\mathbf c$, which is then permuted by a second interleaver $\Pi_2$ to produce the vector $\mathbf s_1$.  The purpose of the shaping encoder is to produce a nonuniform output with a probability $p_0$ of 0 that is higher than the probability $p_1$ of 1.  The detailed operation of the shaping encoder is described in \cite{valenti:2012itc}. The remaining streams at the output of the bit separator are denoted $\{ \mathbf s_2, ..., \mathbf s_5 \}$ and are of the same length as $\mathbf s_1$.  The streams flow into a bit combiner, which passes groups of $m=5$ bits to the APSK modulator.

\begin{figure}
\centering
\vspace{-0.25cm}
\includegraphics[width=3in]{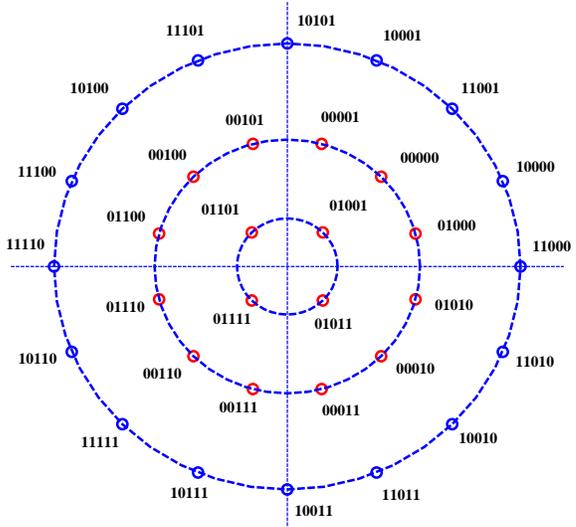}
\caption{32-APSK constellation.  The shaping bit is the MSB.} \label{Fig_APSK}
\vspace{-0.5cm}
\end{figure}

The 32-APSK constellation is shown in Fig. \ref{Fig_APSK} and follows the DVB-S2 standard, except the two most significant bits (MSBs) are exchanged to allow the shaping bit to be the MSB.  The bit from $\mathbf s_1$ is called the {\em shaping} bit and it partitions the constellation into two sub-constellations.  The first partition, which is labeled with an MSB equal to 0, is selected with probability $p_0$, and contains the signals in the inner two rings.  The second partition, which is labeled with an MSB equal to 1, is selected with probability $p_1$, and contains the signals in just the outer ring. Because the shaping code ensures that $p_0 > p_1$, the signals in the inner two rings are more likely to be selected than the high-energy signals of the outer ring.

The overall system rate $R$ is the number of information bits per modulated symbol, and is related to the rate $R_c$ of the LDPC code and the rate $R_s$ of the shaping code by $R = R_c(m + R_s - 1)$.  When uniform modulation is used, $R_s=1$ and there is no shaping code. It is important to note that the choice of $R_c$ and $R_s$ was made to provide a fair comparison between the uniform and shaped system at the same $R$, so the shaped system must use an LDPC code with a higher rate $R_c$ than is used by the uniform system. If the LDPC codes are selected from among those in the DVB-S2 standard (which has just eleven different rates), then the choice of $R_s$ is limited and may be suboptimal. Indeed, one of our motivations for designing new LDPC codes for the shaped system was so that we could select an LDPC code rate $R_c$ that can be matched with a shaping code with a better rate $R_s$.

\section{Baseline Unoptimized Codes}\label{dvbs2}

Before attempting to optimize the LDPC code, we compare the performance of a uniform LDPC-coded system against a comparable shaped system, but restrict the LDPC codes to come from the DVB-S2 standard \cite{dvb:2009} in both cases.  This will serve as a baseline, against which we can compare our optimized codes.   For both shaped and uniform systems, the overall rate is $R=3$ bits/symbol.  To achieve this rate with 32-APSK, the uniform system must use a rate $R_c=3/5$ LDPC code, while for the shaped system, \comment{a higher rate LDPC code is needed to compensate for the rate loss caused by the shaping code.}  a $R_c=2/3$ LDPC code combined with a $R_s =1/2$ shaping code is used for the overall rate to remain $R=3$ bits/symbol. The shaping code with $(n_s,k_s) = (4,2)$ provides a reasonably good $p_0$ while still allowing rates $R_c$ from the DVB-S2 standard to be used (in Section \ref{ldpcoptshape} we will consider custom-rate LDPC codes that will allow a shaping code with better $p_0$ to be used). The codewords of this shaping code are: $\{(0000),(0100),(0010),(0001)\}$.


%

The uniform and shaped systems were simulated over an AWGN channel using long (length $n_c = 64,800$) LDPC codes.  The APSK modulation was configured using {\em ring-radii ratios} that maximize capacity\footnote{Let $\gamma_1$ be the ratio of the radius of the middle ring of the 32-APSK constellation to the radius of the inner ring, and $\gamma_2$ be the ratio of the radius of the outer ring to the radius of the inner ring.  Together, the couple $\{ \gamma_1, \gamma_2 \}$ constitutes the set of {\em ring-radii ratios}.   For both shaped and uniform 32-APSK, the ring-radii ratios that maximize capacity are ($2.64,4.64$)  \cite{valenti:icc2011}}.  The receiver for the uniform system uses the {\em bit-interleaved modulation with iterative decoding} (BICM-ID) architecture described in \cite{xie:vtc2009}\comment{\cite{Caire:1998}}, which involves the iterative exchange of information between the APSK demapper and the LDPC decoder.  A similar BICM-ID receiver was adopted for the shaped system, but with the shaping decoder included in the overall feedback loop.  See \cite{valenti:2012itc} for details of the receiver operation.  For both systems, a maximum of 100 decoder iterations were performed, though the decoder was halted if all the parity-checks were satisfied.

\begin{figure}[t]
\centering
\vspace{-0.1cm}
\includegraphics[width=3.5in,height=2.7in]{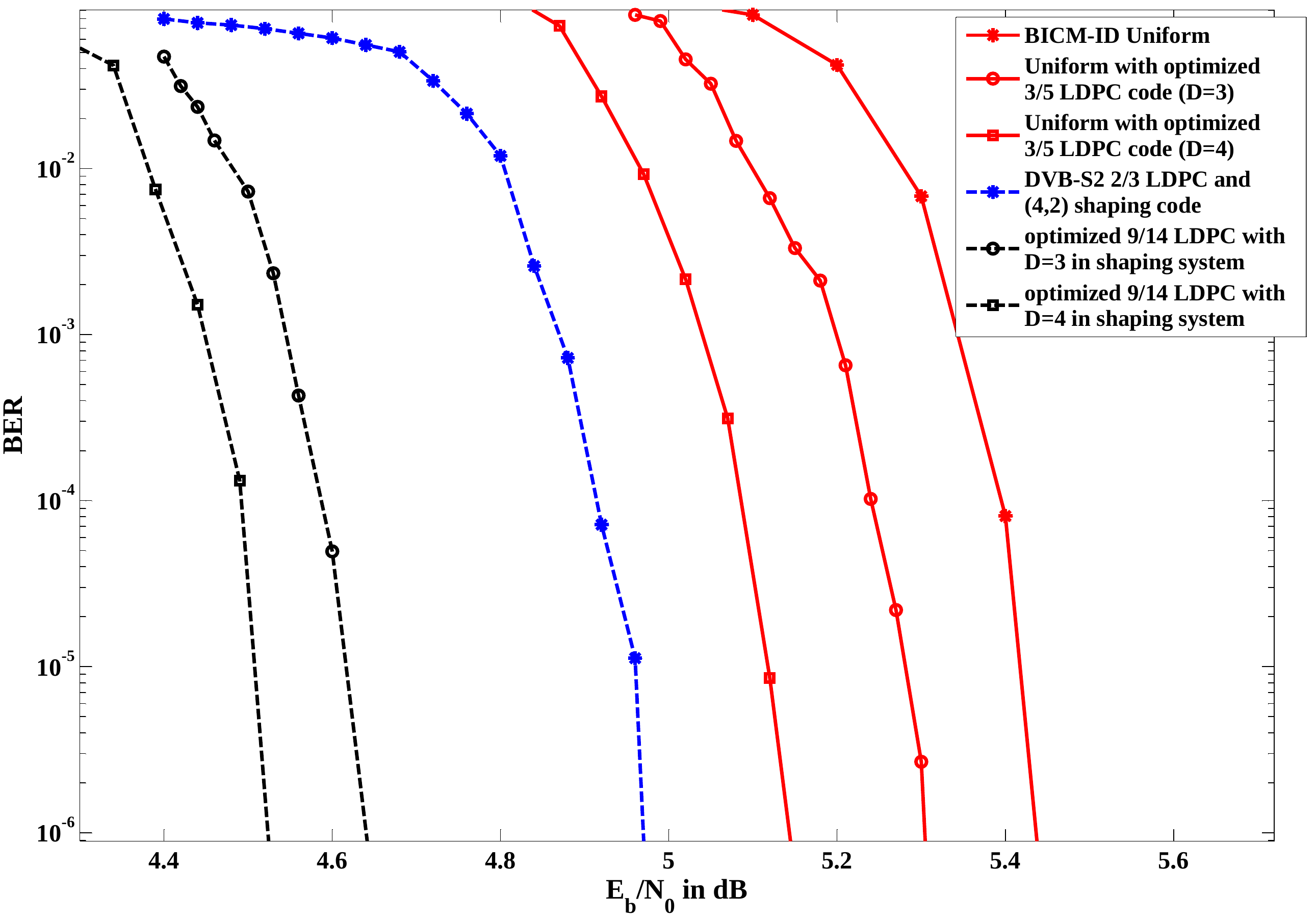}
\vspace{-0.65cm}
\caption{Bit-error rate of 32-APSK in AWGN at rate $R=3$ bits/symbol with a length $n_c = 64\,800$ LDPC code.  The solid lines are for the uniform system; the rightmost curve shows the performance of uniform modulation with a standard DVB-S2 code, while the two curve to its left show the performance of uniform modulation with optimized $D=3$ and $D=4$ LDPC codes, respectively. The dashed lines are for the shaped system; with the left two curves show the performance of shaped modulation with optimized $D=3$ and $D=4$ LDPC codes, and the third curve from left shows the performance of shaped modulation with a standard DVB-S2 code.  } \label{Fig_BER32AWGN}
\vspace{-0.55cm}
\end{figure}

In addition to other curves that will be described later, Fig. \ref{Fig_BER32AWGN} shows the simulated bit-error rates of the two cases described in this section: (1) the uniform system with the rate $R_c = 3/5$ DVB-S2 standardized LDPC code and BICM-ID reception (rightmost curve\comment{; labeled ``BICM-ID uniform'' in the legend}); and (2) the shaped system using the $R_c=2/3$ DVB-S2 standardized LDPC code, $R_s=2/4$ shaping code, and iterative receiver of \cite{valenti:2012itc} (third curve from left). \comment{; labeled ``DVB-S2 2/3 LDPC and (4,2) shaping code} The LDPC codes that are marked ``optimized'' will be described in sections \ref{ldpcopt} and \ref{ldpcoptshape}. From Fig. \ref{Fig_BER32AWGN}, it is observed that the uniform and shaped systems require ${\mathcal E}_b/N_0$ equal to $5.42$ dB and $4.96$ dB to achieve a BER equal to $10^{-5}$, respectively. Thus, the shaped system achieves a gain of $0.46$ dB relative to the uniform system when both use BICM-ID reception.  The capacity limit of 32-APSK with $p_0 = 0.8125$, corresponding to $R_s = 2/4$, is at ${\mathcal E}_b/N_0 = 3.83$ dB \cite{valenti:2012itc}, thus this shaped system is 1.13 dB away from the capacity.

\section{Code Optimization for Uniform Modulation}\label{ldpcopt}
We now optimize the LDPC code.  Our goal is to adapt the EXIT-chart technique of \cite{brink:2004} to account for the behavior of the shaped modulation.  However, before optimizing the LDPC code used by the shaped system, we begin by reviewing how the EXIT charts can be used to optimize the LDPC code in the system with uniform modulation.
To facilitate systematic encoding, we limit the LDPC code to be an extended irregular repeat accumulate (eIRA) code \cite{yang:2004}, which is also the code structure defined in DVB-S2 standard. The parity-check matrix $\mathbf H$ of such a code contains an arbitrary matrix for the first $k_c$ columns and a dual-diagonal matrix as its last $n_c-k_c$ columns. The code is further limited to be {\em check regular}, and its check-node degree $d_c$ is kept the same as the value of the DVB-S2 standardized code of the same rate and length. For instance, the standardized rate $R_c = 3/5$ LDPC code corresponds to the parity-check matrix with dimension $(25 \, 920 \times 64\,800)$ and constant row weight $d_c = 11$, and the comparable optimized code is constrained to have the same check-node degree.

Since the check-node degree is fixed, the optimization problem is to select the optimal variable-node degree distribution. Following \cite{brink:2004}, the variable-node degree is indicated by $d_{v,i}$, $(i=1,...,D)$. The fraction of nodes having degree $d_{v,i}$ is denoted by $a_i$, while the fraction of edges incident to variable nodes of degree $d_{v,i}$ is denoted by $b_i$. First we limit the number of distinct variable-node degrees to be $D=3$, since this is the same number allowed in the DVB-S2 standard. We then increase $D$ by one, and optimize LDPC codes with $D=4$ using the same approach.  Due to the dual-diagonal part of the $\mathbf H$ matrix, the \comment{first} minimum\footnote{In DVB-S2, there is a single degree-1 variable node corresponding to the last column of the $\mathbf H$ matrix.  While our code design maintains this degree-1 node, we do not explicitly list the fraction of nodes connected to degree-1 nodes since it will approach zero for large $n_c$.} variable-node degree is $d_{v,1}=2$.  This implies that the fraction of degree-2 nodes is (at least\footnote{We found no advantage to using $a_1 > (n-k)/n$.}) $a_1=(n-k)/n$; i.e., the fraction of columns corresponding to the dual-diagonal part of $\mathbf H$.
To constrain complexity, the maximum node degree is limited to be no greater than 25; i.e., $d_{v,D} \leq 25$. When $D=3$ is considered, the key step in the optimization is to determine the integer values of $d_{v,2}$ and $d_{v,3}$.  These are related to the constraints on the code rate and total number of \comment{edges} 1s (set by the check-node degree) in the parity-check matrix, thus $d_{v,3}$ can be determined once $d_{v,2}$ is set. Furthermore, once $d_{v,2}$ and $d_{v,3}$ are determined, the corresponding degree distributions, $a_2$ and $a_3$, can be found by using the check-node degree $d_c$.  The optimization is therefore over just one parameter. On the other hand, the optimization is over two parameters when $D=4$; two of the values in the 3-tuple $\{d_{v,2}, d_{v,3}, d_{v,4}\}$ must first be fixed before the remaining value can be determined.





An EXIT chart comprises two EXIT curves \cite{brink:2004}.  One EXIT curve corresponds to the variable-node decoder (VND), and the other one corresponds to the check-node decoder (CND). The EXIT-chart based LDPC code optimization fits the two EXIT curves together by picking appropriate variable-node degrees.  The VND curve characterizes both the variable nodes of the LDPC code and the modulation part.   \comment{The overall VND curve} It is created by first generating a transfer characteristic for just the modulation and its detector at the given  $\mathcal E_s/N_0$. This {\em detector} characteristic is denoted by $I_{E,\mathrm{DET} }( I_{A}, \mathcal E_s/N_0 )$, where $I_{A}$ is computed between the demodulator's {\em a priori} inputs and the modulator's input bits.
For APSK, $I_{E,\mathrm{DET} }$ cannot be generated in closed form for nonzero $I_{A}$, and thus it is generated through Monte Carlo simulation under the assumption that the demodulator's \nopagebreak {\em a priori} input is conditionally Gaussian.

Having found the detector characteristic, the VND curve for degree-$d_v$ nodes is found by using
\begin{multline}\label{VND_COM}
I_{E, \mathrm{VND}} \left(  I_A, d_v, \mathcal E_s/N_0 \right) = \\
J \left( \sqrt{ (d_v-1) [ J^{-1}(I_{A})^2 ] + [J^{-1} (I_{E, \mathrm{DET}}( I_{A}, \mathcal E_s/N_0 ) )]^2 } \right)
\end{multline}
where the $J$-function is given in \cite{brink:2004} and can be computed using the truncated series representation of \cite{torrieri:isrn2011}. Note that (\ref{VND_COM}) only gives the EXIT curve for a single variable-node degree $d_v$, and therefore represents the VND curve for a regular code. The VND curve for an irregular LDPC code is expressed as:
\comment{In the case of an irregular LDPC code, the VND curve is found by using \cite{brink:2004}}
\begin{equation}\label{VND_COM_irregular}
I_{E, \mathrm{VND}} \left(  I_A, \mathcal E_s/N_0 \right)  = \sum_{i=1}^{D} b_i \cdot
I_{E, \mathrm{VND}} \left(  I_A, d_{v,i}, \mathcal E_s/N_0 \right).
\end{equation}
The CND curve is \comment{found by using} given in \cite{brink:2004} as:
\begin{eqnarray}
I_{E, \mathrm{CND}}
\left(  I_A, d_c \right)
& = &
1 - J\left(
\sqrt{d_c-1} \cdot J^{-1}(1-I_A)
\right)
\end{eqnarray}
where $I_A$ is the mutual information at the input of the check nodes.  The EXIT chart is drawn by noting that the $I_{E, \mathrm{CND}}$ produced by the CND becomes the $I_{A}$ at the input to the VND (which we denote $I_{A, \mathrm{VND}}$), while the $I_{E, \mathrm{VND}}$ produced by the VND becomes the $I_{A}$ at the input to the CND (which we denote $I_{A, \mathrm{CND}}$).  The chart plots the VND and CND curves with $I_{A, \mathrm{VND}} = I_{E, \mathrm{CND}}$ on the horizontal axis and $I_{E, \mathrm{VND}} = I_{A, \mathrm{CND}}$ on the vertical axis.  For a given degree distribution, \comment{VND curves are generated at several $\mathcal E_s/N_0$ and} the \emph{threshold} is the value of $\mathcal E_s/N_0$ for which the VND and CND curve just barely touch.

For $D=3$, all feasible combinations of $\{ d_{v,2}, d_{v,3} \}$ were considered and those that lead to low EXIT-chart thresholds were identified.  \comment{For each candidate design, a code with the corresponding degree distribution was realized and simulated.} One code found to be better than rate $R_c=3/5$ DVB-S2 standard code when used with 32-APSK has the following degree distribution:
\begin{center}
\begin{tabular}{ c c c }
  $d_{v,1} = 2$ & $a_1 = 0.40$ & $b_1 =  0.182$\\
  $d_{v,2} = 4$ & $a_2 = 0.52$ & $b_2 =  0.473$\\
  $d_{v,3} = 19$ & $a_3 = 0.08$ & $b_3 =  0.345$\\
\end{tabular}
\end{center}

The process was repeated for $D=4$.  Because the optimization for $D=4$ was over two parameters, there were several degree distributions offering low EXIT-chart thresholds.  A total of 50 candidates were identified, and the realized code for each was simulated.  The code offering the best simulated performance has the following degree distribution:
\begin{center}
\begin{tabular}{ c c c }
  $d_{v,1} = 2$ & $a_1 = 0.40$ & $b_1 =  0.182$\\
  $d_{v,2} = 3$ & $a_2 = 0.10$ & $b_2 =  0.066$\\
  $d_{v,3} = 4$ & $a_3 = 0.44$ & $b_3 =  0.402$\\
  $d_{v,3} = 25$ & $a_3 = 0.06$ & $b_3 =  0.351$\\
\end{tabular}
\end{center}

The uniform 32-APSK system was simulated using a length $n_c = 64,800$ code with these two degree distributions, and the resulting BER curves with BICM-ID reception are shown in Fig. \ref{Fig_BER32AWGN}. The second and third curves from right are generated by using optimized rate $R_c = 3/5$ LDPC code with $D=3$ and $D=4$. These codes indicate the gain of $0.14$ dB and $0.30$ dB relative to the DVB-S2 standard code at a BER equal to $10^{-5}$. The LDPC code with one more variable-node degree ($D=4$) achieves better performance.


\section{Code Optimization for Shaped Modulation}\label{ldpcoptshape}
\begin{figure}[t]
\centering
\vspace{-0.4cm}
\includegraphics[width=3.4in,height=2.4in]{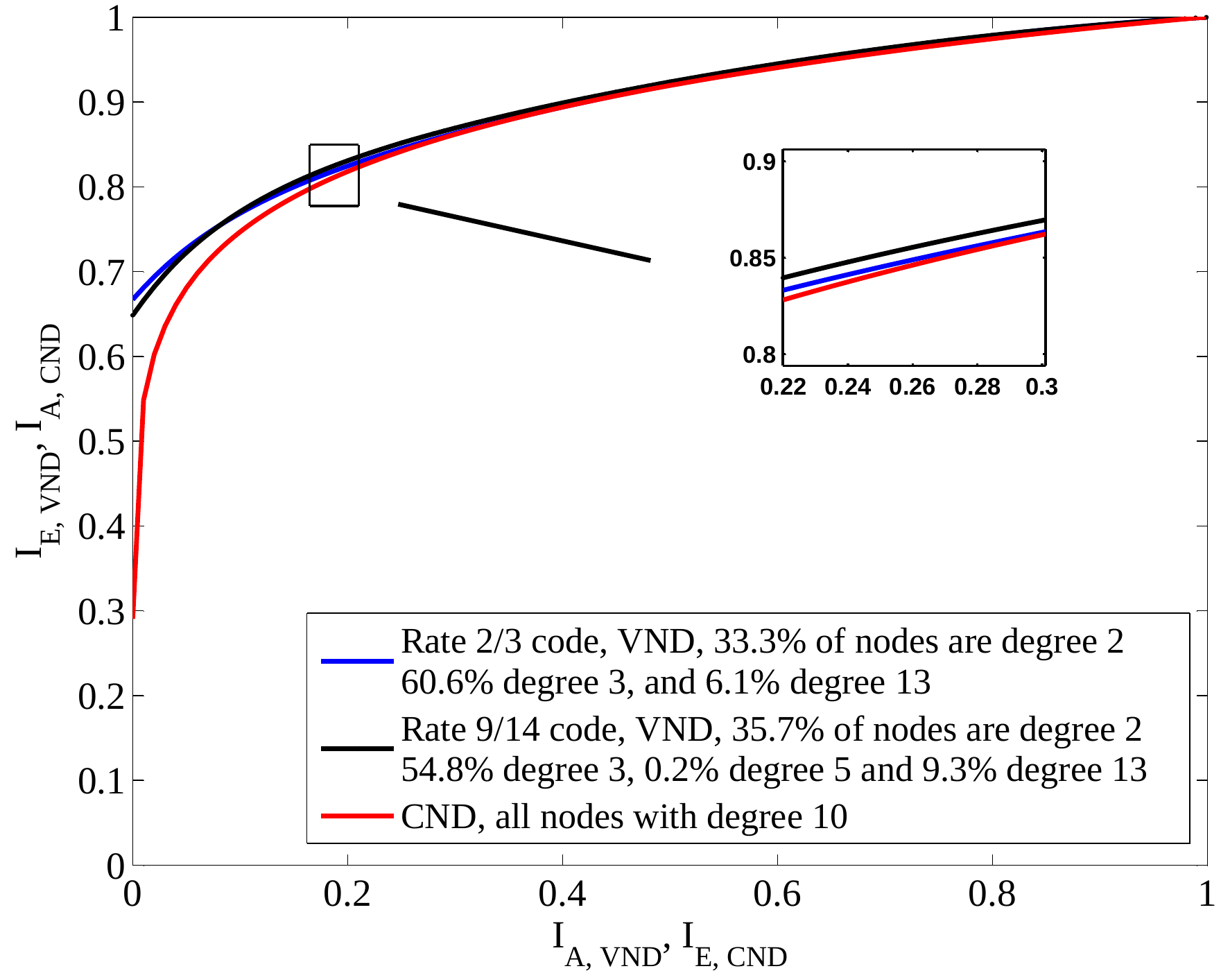}
\vspace{-0.3cm}
\caption{EXIT chart for shaped 32-APSK operating at system rate $R=3$ in AWGN at $\mathcal E_b/N_0 = 4.73$ dB. Two different VND curves are shown, one that is drawn from the DVB-S2 standard and the other is optimized with D=4.} \label{Fig_exit}
\vspace{-0.4cm}
\end{figure}

When shaping is used, its effect is absorbed into the VND curve.
Whereas generating the detector characteristic for the uniform case involves \comment{independently modulating symbols and} measuring the mutual information between the modulator input and demodulator output, the detector characteristic for the \comment{shaped modulation} non-uniform constellation must account for the shaping code.  In reference to Fig. \ref{Fig_transmitter}, a bit sequence $\mathbf v$ is randomly generated and passed through the pictured processing to produce the modulated sequence $\mathbf x$.  The received noisy symbols are processed to produce the extrinsic information $L_e(\mathbf v)$, and the detector characteristic is found by computing the mutual information between $\mathbf v$ and $L_e(\mathbf v)$.
As an example, Fig. \ref{Fig_exit} shows the VND curve of the DVB-S2 standard rate $R_c = 2/3$ LDPC code combined with a ($4,2$) shaping code.  Also shown is the CND curve for check-node degree $d_c = 10$, which is the same $d_c$ of the rate $R_c = 2/3$ LDPC code in the DVB-S2 standard. \comment{which corresponds to the check-node degree of the rate $R_c = 2/3$ LDPC code in the DVB-S2 standard.}

In the previous examples, both the LDPC code rate $R_c$ and the overall rate $R$ are chosen to be compatible with the code rates \comment{found} defined in the DVB-S2 standard.   However, due to the limited number of LDPC code rates, the corresponding \comment{limited} set of possible shaping code rates may \comment{result in the use of a} lead to suboptimal $p_0$.  For instance, the overall system rate $R=3$ bits/symbol can be achieved with an LDPC code rate of $R_c = 2/3$ and a shaping code rate of $R_s = 1/2$.  Using the $(n_s,k_s)=(4,2)$ shaping code will satisfy this requirement, but the resulting value of $p_0$ is $0.8125$, which differs from the optimal value of $p_0=0.716$ identified in \cite{valenti:icc2011}. A $(3,2)$ shaping code, with codewords $\{(000),(100),(010),(001)\}$, gives a $p_0=0.75$, which is closer to the optimal, but requires LDPC code rate $R_c=9/14$ to achieve an overall rate of $R=3$ bits/symbol.

\comment{While the standardized DVB-S2 codes do not support rate $R_c = 9/14$,}
There is no rate $R_c = 9/14$ code defined in DVB-S2 standard.  However,  a new code can be designed for this rate by using EXIT-chart based techniques. As with the rate-2/3 standardized codes, the check node degree of rate $R_c = 9/14$ code is set to be $d_c=10$. \comment{the code has a constant check-node degree of $d_c=10$}  For $D=3$, the code found to have the best performance has variable-node degree distribution:
\begin{center}
\begin{tabular}{ c c c }
  $d_{v,1} = 2$ & $a_1 = 0.357$ & $b_1 =  0.200$\\
  $d_{v,2} = 3$ & $a_2 = 0.558$ & $b_2 =  0.469$\\
  $d_{v,3} = 14$ & $a_3 = 0.085$ & $b_3 =  0.331$\\
\end{tabular}
\end{center}

For $D=4$, the code found to have the best performance has variable-node degree distribution:
\begin{center}
\begin{tabular}{ c c c }
  $d_{v,1} = 2$ & $a_1 = 0.357$ & $b_1 =  0.200$\\
  $d_{v,2} = 3$ & $a_2 = 0.548$ & $b_2 =  0.461$\\
  $d_{v,3} = 5$ & $a_3 = 0.002$ & $b_3 =  0.002$\\
  $d_{v,3} = 13$ & $a_3 = 0.093$ & $b_3 =  0.337$\\
\end{tabular}
\end{center}
The VND curve for this code along with the ($3,2$) shaping code is shown in Fig. \ref{Fig_exit}.  The difference compared with the standard LDPC code is subtle, but we can still observe an open tunnel for the optimized LDPC code.

The BER performance of the two $R_c = 9/14$ length  $n_c = 64\,800$ LDPC codes combined with a $R_s = 2/3$ shaping code  are shown in Fig. \ref{Fig_BER32AWGN}. The leftmost curve is generated by $D=4$ LDPC code, and the curve to its right is generated by using $D=3$ LDPC code.  The results indicate a gain at BER $10^{-5}$ of $0.44$ dB compared to the shaped system that uses the DVB-S2 standard rate $R_c=2/3$ LDPC code along with a $(4,2)$ shaping code, about 0.1 dB of the gain can be attributed to better designed LDPC code with more variable node degree.

\begin{figure}[t]
\centering
\includegraphics[width=3.5in,height=2.5in]{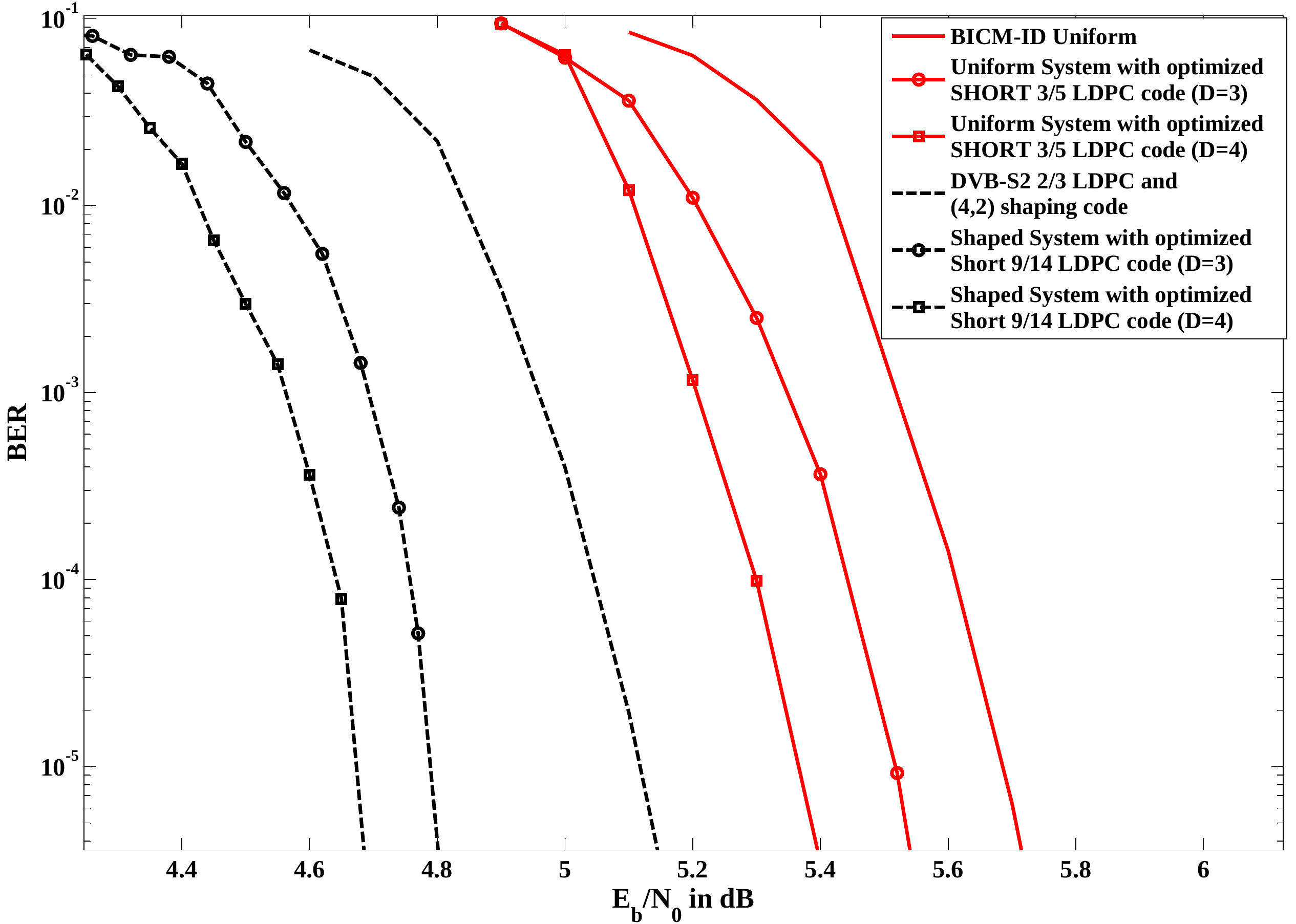}
\vspace{-0.5cm}
\caption{Bit-error rate of 32-APSK in AWGN at rate $R=3$ bits/symbol with short length LDPC codes ($n_c = 16\,200$).  The solid lines are for the uniform system; from right to left, the curves show uniform system using $R_c = 3/5$ DVB-S2 standard code, uniform system with optimized code with $D=3$ and $D=4$. The dashed lines are for the shaped system; from left to right, the curves show performance using optimized $R_c = 9/14$ LDPC code with $D=4$ and $D=3$ combined with $(3,2)$ shaping code, the third curve from left shows the performance of standard $R_c = 2/3$ code with $(4,2)$ shaping code } \label{Fig_BER32AWGNs}
\vspace{-0.5cm}
\end{figure}

\balance

While the results in Fig. \ref{Fig_BER32AWGN} are for long ($n_c = 64\,800$) LDPC codes, Fig. \ref{Fig_BER32AWGNs} compares the bit-error performance of shorter length $n_c = 16\,200$ LDPC codes. In particular, the three solid curves on the right show the performance of uniform systems using short $R_c = 3/5$ LDPC codes, with the second and third curve from right corresponding to the optimized LDPC code with $D=3$ and $D=4$, respectively. The three dashed curves on the left correspond to shaped systems. The third curve from left is generated using the $R_c = 2/3$ LDPC code from DVB-S2 standard combined with a $(n_s, k_s) = (4, 2)$ shaping code. The first and second curve from left incorporate the short $R_c = 9/14$ LDPC code optimized for shaping system with $D=4$ and $D=3$, respectively.  At a BER of $10^{-5}$, optimizing the code for the uniform system provides a gain over the DVB-S2 standardized code of 0.37 dB, while optimizing the code for the shaped system provides a gain of 0.43 dB.

\section{Conclusion} \label{Sec_Conclusion}

Constellation shaping is particularly effective for APSK modulation.  Even if ``off-the-shelf'' codes, such as those in the DVB-S2 standard, are used with shaping, impressive gains can be achieved.  However, an additional and significant gain can be achieved by optimizing the code with respect to the shaped modulation.  In the case of LDPC-coded 32-APSK operating at rate $R=3$, optimizing the degree distribution while allowing $D=4$ distinct variable degrees provides an additional gain of $0.44$ dB for long ($n_c = 64,800$) LDPC codes and $0.43$ dB for short ($n_c = 16,200$) LDPC codes compared to shaped systems that are constrained to use codes from the DVB-S2 standard. If a shaped 32-APSK system with optimized LDPC codes is compared against a uniform system of the same rate ($R=3$), the  shaping gain is between $0.6$ dB and $0.73$ dB, depending on the length of the code and  number of distinct variable node degrees ($D$).  Altogether, the combination of shaping and LDPC degree optimization has the potential to provide a combined (shaping+coding) gain in excess of 1 dB compared to the DVB-S2 standard, which neither shapes the modulation nor optimizes the LDPC degree distribution, and within 0.7 dB of the capacity limit for 32-APSK.

%


\bibliographystyle{ieeetr}
\bibliography{proposal}

\begin{thebibliography}{10}

\bibitem{dvb:2009}
{European Telecommunications Standards Institute}, ``Digital video broadcasting
  {(DVB)} second generation: Framing structure, channel coding and modulation
  systems for broadcasting, interactive services, news gathering and other
  broad band satellite application,'' {\em ETSI EN 302 307 version 1.2.1}, Aug.
  2009.

\bibitem{gaudenzi:2006}
R.~D. Gaudenzi, A.~{Guill\'{e}n i F\`{a}bregas}, and A.~Martinez, ``Turbo-coded
  {APSK} modulations design for satellite broadband communications,'' {\em Int.
  J. Satell. Commun. Network}, vol.~24, pp.~261--281, May. 2006.

\bibitem{valenti:icc2011}
M.~C. Valenti and X.~Xiang, ``Constellation shaping for bit-interleaved coded
  {APSK},'' in {\em Proc. IEEE Int. Conf. on Commun. (ICC)}, (Kyoto, Japan),
  Jun. 2011.

\bibitem{valenti:2012itc}
M.~C. Valenti and X.~Xiang, ``Constellation shaping for bit-interleaved {LDPC}
  coded {APSK},'' {\em IEEE Trans. Commun.}, Oct. 2012.

\bibitem{calderbank:1990}
A.~R. Calderbank and L.~H. Ozarow, ``Nonequiprobable signaling on the
  {Gaussian} channel,'' {\em IEEE Trans. Inform. Theory}, vol.~36,
  pp.~726--740, Jul. 1990.

\bibitem{legoff:2007}
S.~{Le Goff}, B.~K. Khoo, and C.~C. Tsimenidis, ``Constellation shaping for
  bandwidth-efficient turbo-coded modulation with iterative receiver,'' {\em
  IEEE Trans. Wireless Comm.}, vol.~6, pp.~2223--2233, Jun. 2007.

\bibitem{brink:2004}
S.~ten Brink, G.~Kramer, and A.~Ashikhmin, ``Design of low-density parity-check
  codes for modulation and detection,'' {\em IEEE Trans. Commun.}, vol.~52,
  pp.~670--678, Apr. 2004.

\bibitem{xie:vtc2009}
Q.~Xie, K.~Peng, J.~Song, and Z.~Yang, ``Bit-interleaved {LDPC}-coded
  modulation with iterative demapping and decoding,'' in {\em Proc. IEEE Veh.
  Tech. Conf. (VTC)}, (Barcelona, Spain), Apr. 2009.

\bibitem{yang:2004}
M.~Yang, W.~E. Ryan, and Y.~Li, ``Design of efficiently encodable
  moderate-length high-rate irregular {LDPC} codes,'' {\em IEEE Trans.
  Commun.}, vol.~52, pp.~564--571, Apr. 2004.

\bibitem{torrieri:isrn2011}
D.~Torrieri and M.~C. Valenti, ``Rapidly-converging series representations of a
  mutual-information integral,'' {\em ISRN Commun. and Network.}, vol.~2011,
  Article ID 546205, 2011.

\end{thebibliography}

\end{document}